\documentclass[twocolumn,superscriptaddress,aps, prl]{revtex4-2}

\usepackage[dvipsnames]{xcolor}

\usepackage{graphicx}
\usepackage{amssymb}
\usepackage{amsmath}
\usepackage{color}
\usepackage{graphicx}
\usepackage{dcolumn}
\usepackage{bm}
\usepackage{natbib}
\usepackage{epstopdf}
\usepackage[colorlinks=true,citecolor=blue]{hyperref}
\begin{document}
\preprint{APS/123-QED}

\title{Microwave power harvesting using resonator-coupled double quantum dot photodiode}

\newcommand{\orcid}[1]{\href{https://orcid.org/#1}{\includegraphics[width=8pt]{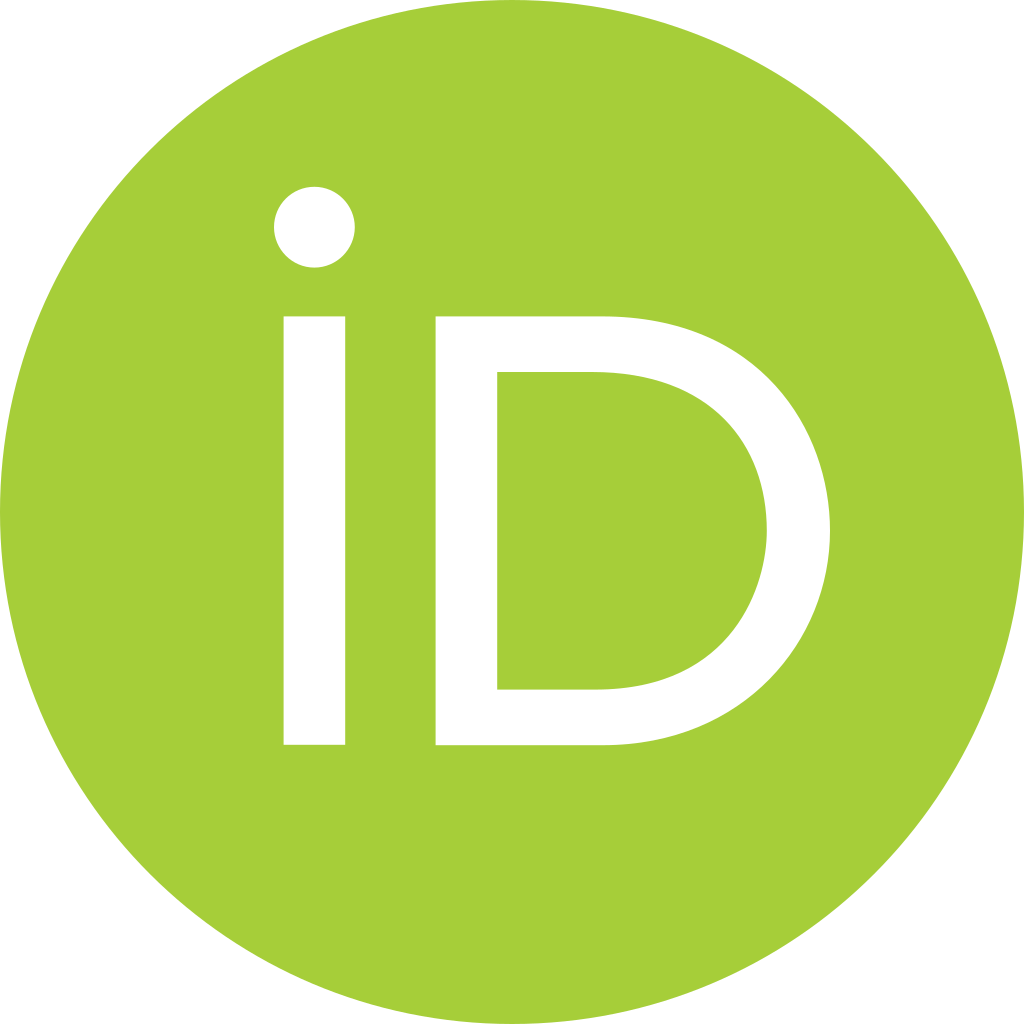}}}
\author{Subhomoy Haldar}
\email{subhomoy.haldar@ftf.lth.se}
\address{NanoLund and Solid State Physics, Lund University, Box 118, 22100 Lund, Sweden}
\author{Drilon Zenelaj} 
 \address{NanoLund and Mathematical Physics, Lund University, Box 118, 22100 Lund, Sweden}
 \author{Patrick P. Potts}
 \address{Department of Physics, University of Basel, Klingelbergstrasse 82, 4056 Basel, Switzerland}
  \author{Harald Havir}
\address{NanoLund and Solid State Physics, Lund University, Box 118, 22100 Lund, Sweden}
\author{Sebastian Lehmann}
\address{NanoLund and Solid State Physics, Lund University, Box 118, 22100 Lund, Sweden}
\author{Kimberly A. Dick}
\address{NanoLund and Solid State Physics, Lund University, Box 118, 22100 Lund, Sweden}
\address{Center for Analysis and Synthesis, Lund University, Box 124, 22100 Lund, Sweden}
 \author{Peter Samuelsson}
 \address{NanoLund and Mathematical Physics, Lund University, Box 118, 22100 Lund, Sweden}
 \author{Ville F. Maisi}
\email{ville.maisi@ftf.lth.se}
 \address{NanoLund and Solid State Physics, Lund University, Box 118, 22100 Lund, Sweden}

\date{\today}

\begin{abstract}
We demonstrate a microwave power-to-electrical energy conversion in a resonator-coupled double quantum dot. The system, operated as a photodiode, converts individual microwave photons to electrons tunneling through the double dot, resulting in an electrical current flowing against the applied voltage bias at input powers down to 1 femto-watt. The device attains a maximum power harvesting efficiency of 2\:\%, with the photon-to-electron conversion efficiency reaching 12\:\% in the single photon absorption regime. We find that the power conversion depends on thermal effects showing that thermodynamics plays a crucial role in the single photon energy conversion.
\end{abstract}

\maketitle

\par Nanoscale energy conversion, enabled by the revolutionary advancements in science and technology~\cite{pelayo2021,zhang2023}, is key to the functionality of solar cells~\cite{krogstrup2013,octavi2011}, telecommunication systems~\cite{portilla2023}, and nanoscale sensing~\cite{wang2012}. Precise control and manipulation of energy at the nanoscale are of fundamental importance to quantum thermodynamics~\cite{pekola2015}, circuit quantum electrodynamics~\cite{kokkoniemi2020,clerk2020}, and quantum information processing~\cite{alexia2022}. For these purposes, quantum dots (QDs) have emerged as compelling candidates, offering a quantum channel to control the energy and information flow~\cite{childress2004, vanderWiel2002, chatterjee2021, barker2022, clerk2020, wang2019}. As demonstrated by previous experiments and theory, QDs have  large potential for thermoelectric power harvesting/heat engines~\cite{josefsson2018, thierschmann2015, bergenfeldt2014}, information-to-work conversion~\cite{barker2022, rafael2019}, and for the conversion of infrared radiation into usable electric power~\cite{sablon2011, supran2015}.

Recently, the investigation of microwave photons interacting with QDs has garnered considerable interest, leading to the development of microwave detectors~\cite{gustavsson2007, wong2017,khan2021,cornia2023}, microwave microscopy~\cite{denisov2022}, and lasing operation in the microwave regime~\cite{liu2014,stockklauser2015,Gullans2015}. These advances provide a unique platform to explore the physics of condensed matter systems and astronomy with an energy resolution of a few tens of micro-electronvolts~\cite{wei2008u, burkard2020,cervantes2022}. Of particular interest is harvesting of microwave power, extensively investigated for large powers ($>$ micro-watts)~\cite{portilla2023,  valenta2014, hemour2014} where a classical wave description of the radiation is appropriate. However,  to date there are no experimental demonstrations of microwave power harvesting in the low power, quantum regime, characterized by single photon absorption, leaving possibilities for utilizing microwave energy at the most fundamental level unexplored.

Here, we present the first experimental realization of a microwave power harvester capable of operating in the single photon absorption regime. An electron occupying a double quantum dot (DQD) embedded in a superconducting resonator absorbs a microwave photon and undergoes photon-assisted tunneling, generating an electrical current from a source (S) to a drain (D) lead. This scheme has previously been demonstrated for microwave photodetectors in Refs.~\cite{wong2017, khan2021}, where S and D are kept at equal chemical potential and the energy gained by the tunneling electrons is dissipated in the leads. In this work, by instead operating the photodetector against an applied bias, part of the energy gained by absorbing a microwave photon is converted into electrical energy. We hence realize microwave power harvesting in the quantum regime, observing a maximal power harvesting efficiency of 2\:\% at 1 femto-watt input power. Our results open up for applications in the single-photon absorption regime, in on-chip power delivery in quantum circuits ~\cite{alexia2022, clerk2020} and astronomy, enabling energy collection from cosmic electromagnetic radiation~\cite{wei2008u, cervantes2022, haipeng2023}.
\begin{figure}
\includegraphics[width=3.3in]{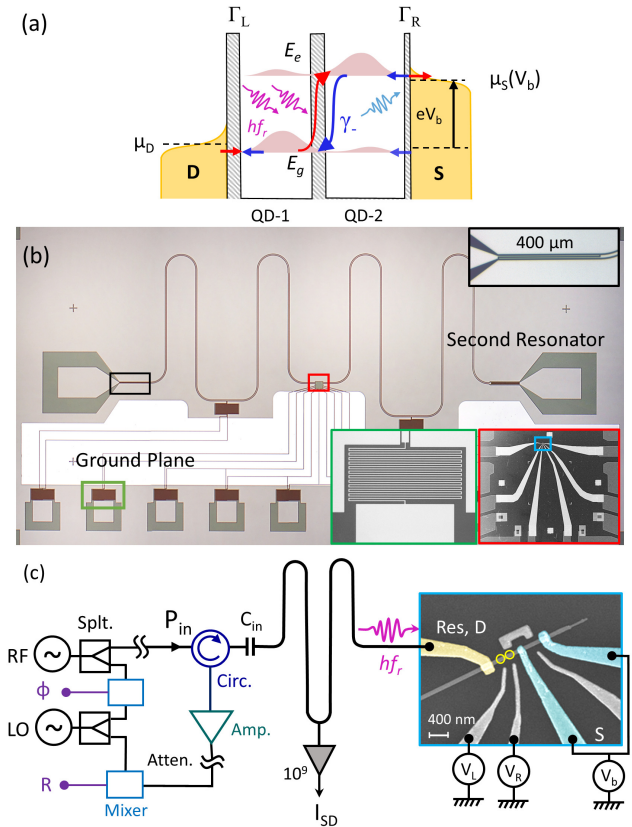}
\caption{\label{fig1}(a) Schematic of device energy band diagram. Red arrows show an electron absorbing a photon of energy $hf_r$ and tunneling through the DQD against a bias voltage $V_b$. Blue arrows show tunneling processes in the direction of applied bias. (b) Optical micrograph of the hybrid device. Insets depict the input coupler of the resonator, inductive filtering on the DC lines, and the central part of the device showing the connections to the DQD. (c) False-colored scanning electron micrograph showing the nanowire device with the schematic layout of the RF and transport measurement scheme. Bias is applied on the blue leads across the DQD (indicated by yellow circles) and current $I_\text{SD}$ is measured via the middle point of the resonator that directly connects the DQD (yellow lead). 
}
\end{figure}

The device operation is illustrated schematically in Fig.~\ref{fig1}a. In the single photon absorption regime, under ideal operation conditions \cite{khan2021} every single photon is converted into an electron tunneling through the DQD (red arrows), against the bias. At low temperatures, for a bias approaching the photon energy, the efficiency of the power conversion as well as the photon-to-electron conversion, or photodetection, efficiency thus reach unity. For an increasing temperature,  thermally induced excitations in the DQD block the photon-assisted tunneling and lead to a backflow of electrons (blue arrows), reducing both the photodetection and power conversion efficiencies. We here present a theoretical estimate of the thermodynamic maximum of the power conversion efficiency. Moreover, increasing the power beyond the single photon absorption regime reduces the photodetection and power conversion efficiencies as well. Our scheme is qualitatively different from conventional microwave power conversion with rectifier-antennas, or rectennas ~\cite{joshi2013, song2015, donchev2014, ladan2013}. Rectennas work at high powers, where the radiation can be described as classical waves. The maximum power conversion efficiency, which can approach unity, is set by the device properties and not temperature - operation is typically at ambient temperatures.

\begin{figure}
\includegraphics[width=3.4in]{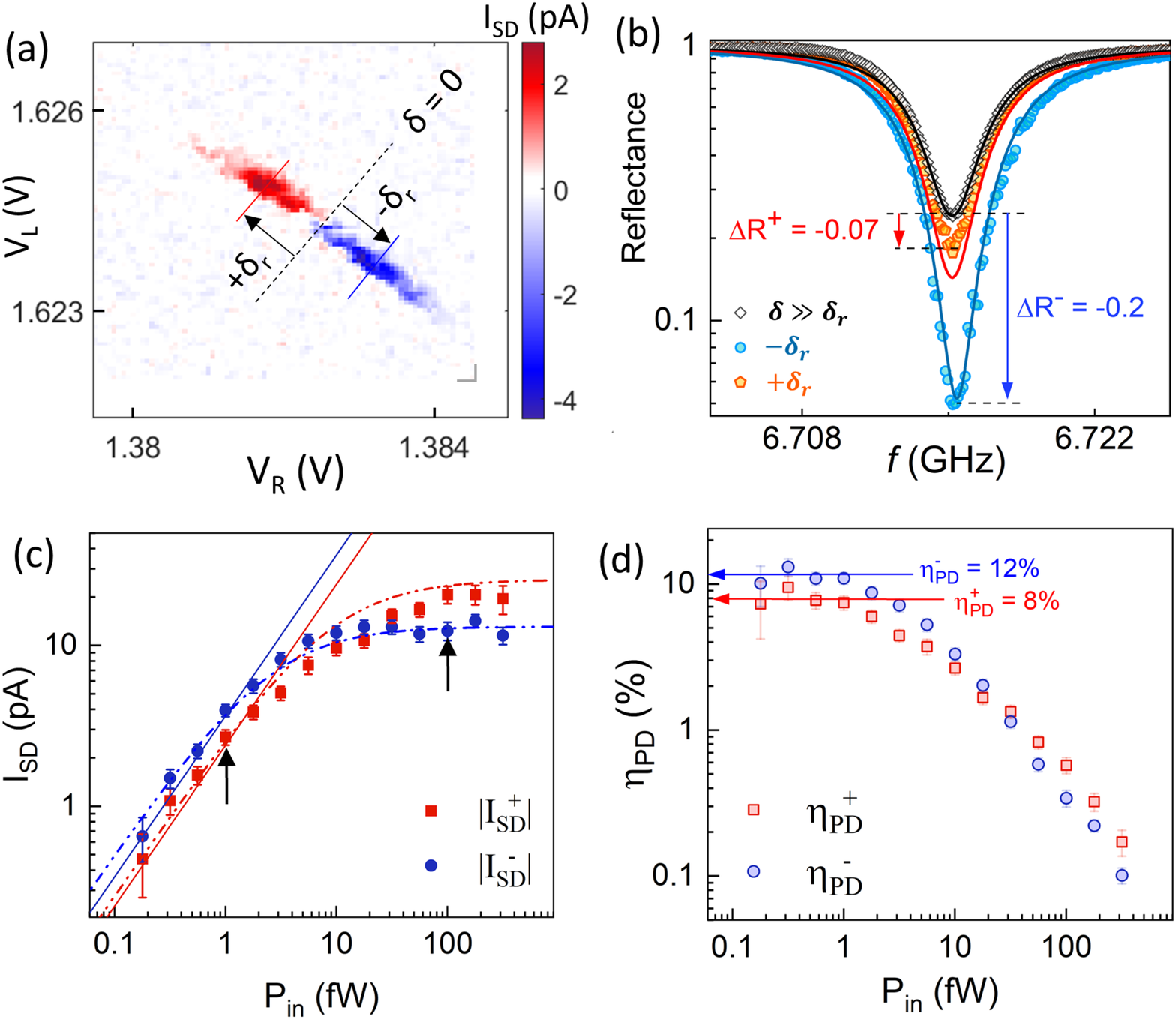}
\caption{\label{fig2}(a) Measured photocurrent, $I_\mathrm{SD}$ as a function of gate voltages $V_L$ and $V_R$ with $V_b=0$ and microwave drive $P_\mathrm{in}$ = 1~fW. The black dashed line indicates the $\delta$ = 0 detuning line. (b) Microwave reflection coefficient as a function of drive frequency measured in the Coulomb blockade regime (at $\delta\gg\delta_r$) and photo-detection point (at $\delta=\pm~\delta_r$) with $P_\mathrm{in}$ = 1~fW. (c) Photocurrent and (d) photon-to-electron conversion efficiency as a function of $P_\mathrm{in}$. Solid (dashed) lines show the theoretically fitted curves at the low (large) drive regime.}
\end{figure}

The device comprises a semiconductor DQD dipole coupled to a one-port waveguide resonator, Fig.~\ref{fig1}b. The DQD is formed by three segments of wurtzite barriers that confine electrons in two zincblende islands in an InAs nanowire~\cite{Luna2015,barker2019, thelander2011}. Two plunger gate voltages $V_\mathrm{L}$ and $V_\mathrm{R}$ control the electrochemical potential of QD-1 and 2, respectively. The D-lead of the DQD directly connects to the resonator which has a fundamental resonance at $f_r$ = 6.715 GHz. The current $I_\mathrm{SD}$ across the DQD is measured via the resonator voltage node point. The nanowire also hosts a third (sensor) dot coupled to a readout resonator with fundamental resonance at 6.31 GHz ($\neq$ $f_r$), see Fig.~\ref{fig1}b. We do not use the sensor dot or the second resonator in the present study. To avoid any influence from the third dot, we apply the same bias $V_b$ to both sides of the sensor dot, Fig.~\ref{fig1}c. Measurements are performed in a dilution refrigerator with an electronic temperature of $T_e$ $\approx$ 60 mK. The supplementary information (SI)~\cite{supp} presents the DC current-bias measurements of the DQD and details of the rf heterodyne circuit used in this work. The same device has been used in Ref.~\cite{haldar2022} to investigate wave-particle interplay by studying the energetics of the microwaves during the photodiode operation. For the power conversion, we drive the system with the cavity resonance frequency $f_r$, and only use $f\neq f_r$ for the device characterization.

\begin{figure}
   \centering
  \includegraphics[width=3.4in]{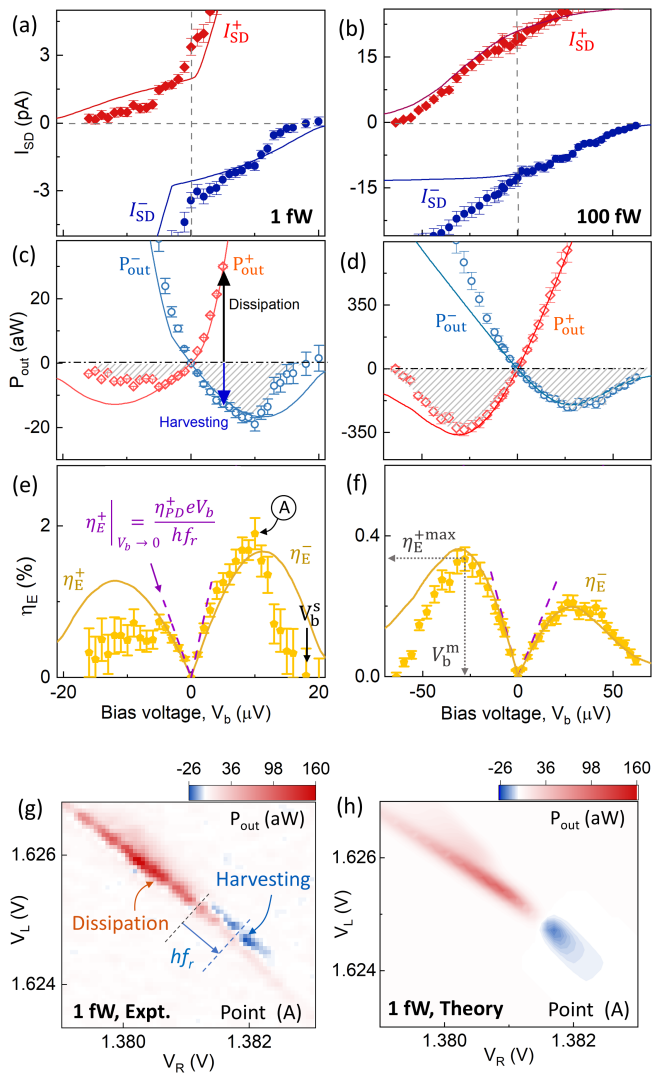}
 \caption{Photocurrent measured with an applied bias $V_b$ across the DQD for (a) 1~fW and (b) 100~fW input microwave powers. The solid lines depict the numerically calculated results for the two detuning configurations. Panels (c-d) show the corresponding power harvested or dissipated by the device, while panels (e-f) show the corresponding power harvesting efficiency $\eta_E$ for the two input powers. The bias setting required to achieve maximum $\eta_E$ of 2\% is marked with point (A) in figure (e). Panels (g-h) present experimental and theoretical results comparing the power harvested and dissipated in the plunger gate space with $P_\mathrm{in}$ = 1~fW and $V_b$ = 10~$\mu$V, which corresponds to point (A) in figure (e).}
\label{fig3}
\end{figure}

We start by measuring the photo-response without applying a bias voltage. Figure~\ref{fig2}a shows the photocurrent as a function of gate voltages with microwave power $P_\mathrm{in}$ = 1~fW. By varying $V_\text{L}$ and $V_\text{R}$, we move the energy levels of the respective dots and thus change the detuning, $\delta$. In Fig.~\ref{fig2}a, we observe that $I_\text{SD}^\pm$ in the direction of $\pm\delta$ reaches a maximum when the energy gap between the hybridized states of the DQD matches the photon energy, $hf_r=E_e-E_g=\sqrt{\delta_r^2+4t^2}$, where $E_g$ ($E_e$) denotes the energy of the ground (excited) state and \textit{t} denotes the interdot tunnel coupling~\cite{vanderWiel2002, khan2021}. Due to the energy transfer from the photons to the electrons, the DQD at $\pm\delta_r$ introduces an additional loss channel ($\kappa_\text{DQD}^\pm$) to the resonator photon mode. Figure~\ref{fig2}b presents the changes to the resonator response in the different detuning conditions due to this additional loss channel. We then vary $P_\mathrm{in}$ and plot the measured $I_\text{SD}$ at $\pm\delta_r$ in Fig.~\ref{fig2}c. The corresponding photon-to-electron conversion efficiency, $\eta_\mathrm{PD}=I_\text{SD} hf_r/eP_\text{in}$, is plotted in Fig.~\ref{fig2}d, which shows a plateau in the low-power regime with $\eta_\text{PD}^-$ reaching 12\%. Note that $\eta_\text{PD}$ encompasses the entire process of photon-assisted tunneling, and thus corresponds to the external quantum efficiency (see the SI~\cite{supp}).

Next, to explore the ability of our device as a microwave power harvester, we consider $P_\mathrm{in}$ = 1 fW and 100 fW. This represents two distinct physical regimes, with linear and non-linear response to the photon flux, see the marked arrows in Fig.~\ref{fig2}c. Figures~\ref{fig3}a and \ref{fig3}b present the maximum $I_\text{SD}^\pm$ in the plunger gate space recorded as a function of $V_b$. The corresponding electrical power out $P_\mathrm{out} = I_\mathrm{SD}V_b$ is shown in Figs.~\ref{fig3}c and \ref{fig3}d. For a wide range of bias voltages, $I_\mathrm{SD}$ flows against $V_b$ and hence, microwave energy is converted into electrical energy. The shaded areas indicate the overall bias range where power harvesting takes place. For opposite detuning, $V_b$ drives the electronic transport in the bias direction, resulting in a sharp increase in the power dissipation. The blue (black) arrow of Fig.~\ref{fig3}c depicts the amount of power harvested (dissipated) with $V_b$ = 5~$\mu$V when the energy levels are detuned to $-\delta_r$ ($+\delta_r$). 

We employ the commonly used ~\cite{song2015, joshi2013} definition of the power harvesting efficiency as the ratio of the DC electric output power to the AC input power from the microwave drive, as
\begin{equation}
    \eta_E=\frac{P_\mathrm{out}}{P_\mathrm{in}}=\frac{I_\mathrm{SD}V_b}{P_\mathrm{in}}.
\label{etaEeq}
\end{equation}
Figures~\ref{fig3}e and \ref{fig3}f plot $\eta_E$ as a function of $V_b$ for the two input powers. For $eV_b\ll hf_r$, the $\eta_E$ increases linearly with $V_b$. In this low-bias regime, the bias-independent nature of $\eta_\text{PD}$ allows us to observe a linear dependence of $\eta_\text{E}^\pm=\eta_\text{PD}^\pm eV_b/hf_r$, as illustrated by the dashed purple lines in Figs.~\ref{fig3}e and \ref{fig3}f. Further increasing $V_b$ causes a sub-linear response in $\eta_E$, which then reaches a maximum $\eta_E^\text{max}$ with $V_b^m$, followed by a sharp decrease when $eV_b$ approaches the photon energy $hf_r$. For $P_\mathrm{in}$ = 1~fW, the device attains the maximum power harvesting efficiency $\eta_E^\text{-max}$ = 2\% with positive bias $V_b^m$ = 10~$\mu$V as marked by (A) in Fig.~\ref{fig3}e. 

The energy of the absorbed radiation contributing to the power harvesting is set by the DQD level detuning $\delta$. For the low-power case, with $P_\text{in}$ = 1~fW, it is clear from Fig.~\ref{fig3}g (see also Fig. ~\ref{fig2}a) that power is harvested at the level detuning $\delta \approx hf_r$. For $\delta \gg hf_r$ the photocurrent and hence the power harvesting becomes vanishingly small. This shows that the power harvesting occurs in the single photon absorption regime, with negligible contributions from multiple-photon absorption. In line with this observation, the power harvesting efficiency in Fig.~\ref{fig3}e vanishes when the applied bias $eV_b$ exceeds the single photon energy $hf_r$. On the contrary, for $P_\text{in}$ = 100~fW, power harvesting occurs even when $eV_b > hf_r$, as seen in Fig.~\ref{fig3}f. This can mainly be attributed to the non-linear, multiple photon absorption process. Moreover, the broadening of the photocurrent spot around $\delta_r$ with strong drive also enables power harvesting at $eV_b > hf_r$.

The effect of temperature on the power conversion efficiency is of particular interest in our device, where the thermal energy ($k_\text BT_\text e$) of the electrons cannot be neglected compared to the photon energy, $hf_r$. For an applied bias voltage such that the chemical potentials of the drain and source leads approach the energy levels ($E_g, E_e$) of the DQD, thermal fluctuations in the electron population near the Fermi levels of the reservoirs induce excitations to the DQD occupancy. As shown in Fig.~\ref{fig1}a, this hinders the photon absorption in the DQD (red arrows) and also leads to a backflow of the electrons (blue arrows), suppressing the power conversion efficiency. To obtain a qualitative understanding of these thermodynamic effects on the power harvesting, we consider a basic, mean-field master equation model (see SI for details~\cite{supp}), assuming low drive power, symmetric tunnel couplings, no interdot relaxation and no internal losses in the resonator. At $T_e=0$, these asumptions give $\eta_E=1$ as well as $\eta_\text{PD}=1$. At non zero temperatures, $k_BT_e\ll hf_r$, the power conversion efficiency in Eq. (\ref{etaEeq}), maximized over bias voltage, can then conveniently be written
\begin{equation}\label{eq3}
\eta_E^\text{max}=\frac{eV_b^\text m}{hf_r}\eta_\text{PD}^\text{max}, 
\end{equation}

where $V_b^\text m=hf_r-2k_BT_e\ln(1/2\sqrt{hf_r/k_BT_e}+9/4)$ is the maximizing bias voltage and $\eta_\text{PD}^\text{max} = (1 - k_B T_e/hf_r)$ is the photo-detection efficiency attainable at $V_b^\text m$. Equation~(\ref{eq3}) shows that two separate thermal effects suppress the maximum power harvesting efficiency, $\eta_E^\text{max}$. First, the photo-detection efficiency $\eta_\text{PD}$ is suppressed by increasing temperature. Second, the maximizing bias voltage $V_b^\text m$ decreases to a value below the photon energy $hf_r$. As a result, we have $\eta_E^\text{max}\leq \eta_\text{PD}^\text{max}$. Moreover, since $\eta_\text{PD}^\text{max}$ is the photo-detection efficiency at $V_b^\text m$ and not at zero bias (for which it is maximal), it is clear that temperature effects have a more detrimental influence on $\eta_E^\text{max}$ than on the maximal $\eta_\text{PD}$ . This is in agreement with our findings that $\eta_\text{E}^\text{max}=2$ \% while the maximal, low power, low bias $\eta_\text{PD}$ reaches 12\%.

The device properties are obtained by comparing the zero applied bias photocurrent data of Fig.~\ref{fig2}c and the resonator response in Fig.~\ref{fig2}b with theoretical predictions. We use the quantum master equation approach of Khan \textit{et al.}~\cite{khan2021} with the additional features of $I_\text{SD}^+\neq I_\text{SD}^-$ which allow us to determine the tunnel rates at the left ($\Gamma_\text{L}$) and right ($\Gamma_\text{R}$) QD-lead barriers individually. We calculate six unknown parameters: tunnel rates $\Gamma_\text{L}$, $\Gamma_\text{R}$, interdot tunnel coupling $t$, charge dephasing rate $\gamma_\phi$, interdot relaxation rate $\gamma_-$, and cavity-DQD coupling constant \textit{g}, by fitting the following six features: two photocurrent slopes in the low-power limit of Fig.~\ref{fig2}c, two photocurrent plateaus in the high-power limit of Fig.~\ref{fig2}c, and the rf reflectance curves measured at $\pm\delta_r$ in Fig.~\ref{fig2}b. We note that $t$ contributes to the hybridization of the DQD levels and that the coupling constant $g$ sets the interaction strength between the resonator and DQD, determining the response of the DQD to the drive of the microwave cavity ~\cite{khan2021, wong2017, burkard2020}. The solid (dashed) lines in Fig.~\ref{fig2}c show the fitted curves in the linear (non-linear) response regime using Eq.~(S.6) and Eq.~(S.9), see the SI for further details~\cite{supp}. We see that the theoretical results capture the essential features of the experimental data, with the best fitting yielding $\Gamma_{\text{L}}/2\pi$ = 27 MHz, $\Gamma_{\text{R}}/2\pi$ = 5400 MHz, $\gamma_-/2\pi$ = 36 MHz, tunnel coupling $t$ = 2.5~$\mu$eV, $\gamma_\phi/2\pi$ = 1500 MHz and $g/2\pi$ = 27 MHz. Importantly, we find $\Gamma_{\text{R}}\gg$ $\Gamma_{\text{L}}, \gamma_-$, which is causing the asymmetry in the device operation. The fits (Eq.~S.14) to the reflectance spectra in Fig.~\ref{fig2}b provide the couplings, $\kappa_\mathrm{DQD}^{+}/2\pi$ = $4g^2/2\pi\Tilde{\Gamma}^+$ = 0.4~MHz and $\kappa_\mathrm{DQD}^{-}/2\pi$ = $4g^2/2\pi\Tilde{\Gamma}^+$ = 1.0~MHz, where $\Tilde{\Gamma}^+$ ($\Tilde{\Gamma}^-$) is the total decoherence rate at $+\delta_r$ ($-\delta_r$). Based on the scatter of the data points and their correspondence to the fits, we estimate an uncertainty of approximately 10\:\% for all the parameter values.

To get further insight into the asymmetry of the device operation, we derive an expression for the photocurrent in zero-bias setting for low drive power and $\Gamma_{\text 0e}\gg \kappa$, reading
\begin{equation}\label{efficiency}
   I_\text{SD}/e=\frac{P_\text{in}}{hf_r}\frac{\kappa_\text{c}}{\kappa}\frac{4\kappa_\text{DQD}\kappa}{(\kappa_\text{DQD}+\kappa)^2}\frac{\Gamma_{\text{0}e}}{(\Gamma_{\text{0}e}+\gamma_-)}p_f,
\end{equation}
where $\Gamma_{g0}=\Gamma_{g\text{L}}+\Gamma_{g\text{R}}$ ($\Gamma_{0e}=\Gamma_{\text{L}e}+\Gamma_{\text{R}e}$) is the total tunnel rate into the ground (out of the excited) state and $p_f$ = $(\Gamma_{g\text{L}}\Gamma_{\text{R}e}-\Gamma_{g\text{R}}\Gamma_{\text{L}e})/\Gamma_{0e}\Gamma_{g0}$ denotes the directivity of the DQD. Here $\Gamma_{(\text{L/R})(g/e)}$ ($\Gamma_{(g/e)(\text{L/R})}$) denotes the tunneling rates out from (in to) the DQD ground/excited state from the left/right leads. These tunnel rates are the couplings $\Gamma_\text{L/R}$ weighted with the corresponding hybridization of the DQD state. The asymmetry of the photo-response can be traced back to the asymmetry in $\kappa_\mathrm{DQD}$. At $+\delta_r$, the tunnel rate out of the excited state is much larger than at $-\delta_r$ due to $\Gamma_{\text{R}}\gg\Gamma_{\text{L}}$. This results in a larger decoherence rate $\tilde{\Gamma}$ at $+\delta_r$, which reduces the effective coupling  $\kappa_\mathrm{DQD}$ in analogy to the Zeno effect~\cite{hackenbroich1998}. In addition, the fourth term in Eq.~\eqref{efficiency} adds a smaller contribution to the asymmetry. 

In the high power regime [Eq.~(S.9)], the ground and excited states become equally populated, and thus the dynamics within the DQD can be neglected. Here, the saturation photocurrent only depends on the in- and out-tunneling rates and is given by $I_\text{SD}^{*}/e=p_f/(1/\Gamma_{g0}+2/\Gamma_{0e})$, where the factor of 2 arises from the fact that the DQD spends only half of the time in the excited state. In the limit of large asymmetry $\Gamma_\text{L}\ll\Gamma_\text{R}$, the smallest rate among $\Gamma_{g0}$ and $\Gamma_{0e}$  determines the $I_\text{SD}^*$ and the ratio of currents becomes $I_\text{SD}^\text{*+}/I_\text{SD}^\text{*-}=2\Gamma_{\text{L}e}^+/\Gamma_{g\text{L}}^-= 2$, which is in good agreement with our experiment $I_\text{SD}^\text{*+}/I_\text{SD}^\text{*-}$ = 1.7. 

Based on the parameter values extracted from the zero bias data, we can numerically evaluate the finite bias current $I_\text{SD}^\pm$ and the corresponding power conversion at $T_e$ = 60~mK. The solid lines in Figs.~\ref{fig3}(a-f) depict the results. Furthermore, in Figs.~\ref{fig3}g and \ref{fig3}h, we compare the experiment and theory results showing the amount of power harvested and dissipated by the device in the plunger gate space corresponding to the point (A) of Fig.~\ref{fig3}e. Once again, the theoretical curves are found to be in good agreement with the experimental data and reproduce the asymmetric $\eta_E$ in the two detuning configurations. We however note that for chemical potentials close to the DQD energies our master equation approach formally breaks down, which may explain the discrepancy between experiment and theory at $|V_b| \sim hf_r$ in Fig.~\ref{fig3}(c) and (e). Moreover, despite that our theoretical approach does not account for multi-photon processes, the agreement between experiment and theory for $P_{\text in}=100$~fW in Fig.~\ref{fig3}(c) and (e) is still good, see SI for further discussion~\cite{supp}.

In conclusion, we have demonstrated a microwave resonator-DQD power harvester operating in the femtowatt power, single photon absorption regime where a quantum description of the incoming radiation is appropriate and conventional rectenna power harvesters do not work. We achieve a maximum power conversion efficiency of 2\:\% and a photon-to-electron conversion efficiency reaching 12\:\%.  Our controllable and versatile microwave power harvester opens up avenues of experiments enabling energy collection at the single-photon absorption limit, to explore fundamental aspects of quantum thermodynamics, quantum optics, and astrophysics.

\begin{acknowledgments}
We acknowledge fruitful discussions with Antti Ranni, Claes Thelander, Adam Burke, Martin Leijnse, and Andreas Wacker and the financial support from the Knut and Alice Wallenberg Foundation through the Wallenberg Center for Quantum Technology (WACQT), the Foundational Questions Institute, a donor advised fund of Silicon Valley Community Foundation (grant number FQXi-IAF19-07), NanoLund, and Swedish Research Council (Dnr 2019-04111). We also thank Jonas Bylander, Simone Gasparinetti, and Anita Fadavi Roudsari from the Chalmers University of Technology for fruitful discussion to improve the superconducting cavity. P.P.P. acknowledges funding from the Swiss National Science Foundation (Eccellenza Professorial Fellowship PCEFP2\_194268).
\end{acknowledgments}

\bibliography{master}
\end{document}


\renewcommand{\theequation}{S.\arabic{equation}}
\renewcommand{\thefigure}{S.\arabic{figure}}

\newcommand{\ville}[1]{{\color{magenta} #1}}

\title{Supplementary Information: Microwave power harvesting using resonator-coupled double quantum dot photodiode}

\newcommand{\orcid}[1]{\href{https://orcid.org/#1}{\includegraphics[width=8pt]{orcid.png}}}
\author{Subhomoy Haldar}
\email{subhomoy.haldar@ftf.lth.se}
 \address{NanoLund and Solid State Physics, Lund University, Box 118, 22100 Lund, Sweden}
 \author{Drilon Zenelaj}
 \address{Physics Department and NanoLund, Lund University, Box 118, 22100 Lund, Sweden}
 \author{Patrick P. Potts}
 \address{Department of Physics, University of Basel, Klingelbergstrasse 82, 4056 Basel, Switzerland}
  \author{Harald Havir}
 \address{NanoLund and Solid State Physics, Lund University, Box 118, 22100 Lund, Sweden}
 \author{Sebastian Lehmann}
 \address{NanoLund and Solid State Physics, Lund University, Box 118, 22100 Lund, Sweden}
\author{Kimberly A. Dick}
 \address{NanoLund and Solid State Physics, Lund University, Box 118, 22100 Lund, Sweden}
 \address{Center for Analysis and Synthesis, Lund University, Box 124, 22100 Lund, Sweden}
 \author{Peter Samuelsson }
 \address{Physics Department and NanoLund, Lund University, Box 118, 22100 Lund, Sweden}
 \author{Ville F. Maisi}
 \email{ville.maisi@ftf.lth.se}
 \address{NanoLund and Solid State Physics, Lund University, Box 118, 22100 Lund, Sweden}
\maketitle
\onecolumngrid
Further information is provided here on the DC transport measurements, details of our rf heterodyne circuit to measure the photo-response and internal/external quantum efficiency of the device. We also present the theoretical calculations and expressions used in the modelling of the microwave reflectometry and charge transport measurement presented in the main text. Moreover, we discuss the effect of temperature on the power harvesting and photodetection efficiencies.

\section{DC Transport measurements}

\begin{figure}[h]
\includegraphics[width=5.7in]{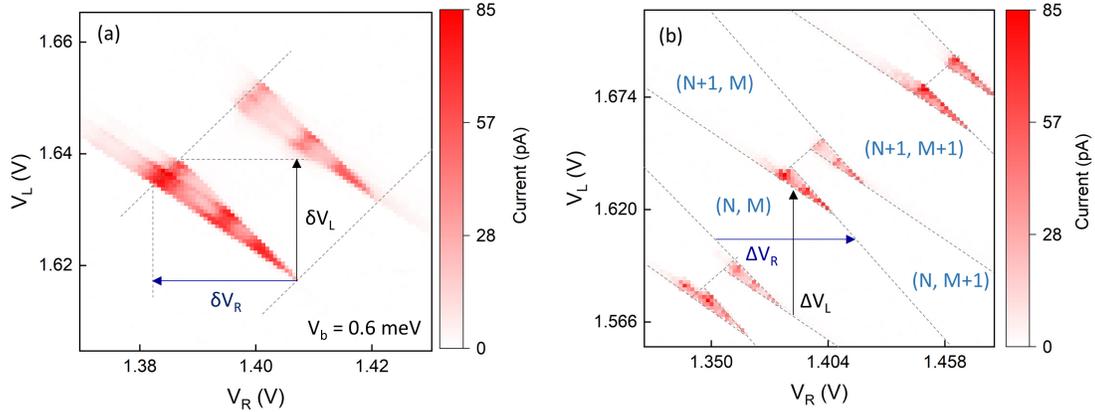}
\caption{\label{fig1s}(a): Charge transport measurements of the DQD with $V_b$ = 600 $\mu$V as a function of plunger gate voltages $V_L$ and $V_R$. Dotted lines are a guide to the eye. (b): Bias-driven ($V_b$ = 600~$\mu$V) current across the DQD in a larger plunger gate space. The electron occupancy (N, M) in the DQD remains the same within a hexagonal charge stability pattern. Thus, the energy cost to add an electron to the left (right) dot is given by $\alpha_L\Delta V_L$ ($\alpha_R\Delta V_R$).}
\end{figure}

Supplementary Fig.~\ref{fig1s} presents the standard DC transport measurements of the nanowire DQD without applying a microwave drive. With the source-drain bias $V_b$ = 600 $\mu$V, applied across the DQD, tunneling of electrons is allowed. We observe distinct features inside the finite bias triangles (FBTs) arising due to the charge transport via the excited states of the DQD. The energy levels of the respective QDs linearly change with the gate voltages with $\Delta E_\text{L} = \alpha_\text{L}~\delta V_\text{L}$ and $\Delta E_\text{R} = \alpha_\text{R}~\delta V_\text{R}$, where the proportionality constant is known as the lever arm. From the FBT of Fig.~\ref{fig1s}(a), we determine the lever arms $\alpha_\text{L}$ = 27~$\mu$eV/mV and $\alpha_\text{R}$ = 24~$\mu$eV/mV~\cite{vanderWiel2002, taubert2011}. Figure~\ref{fig1s}(b) shows the transport measurements in a bigger plunger gate space. As a result, two new sets of FBTs are also visible in this case. We use Fig.~\ref{fig1s}(b) to determine the charging energy of the dots, $E_\text{c,L}$ = $\alpha_\text{L}\Delta V_\text{L}$ = 1.7~meV and $E_\text{c,L}$ = $\alpha_\text{L}\Delta V_\text{L}$ = 1.6~meV. 

\section{Details of the RF measurements}
\begin{figure}[h]
\includegraphics[width=6.0in]{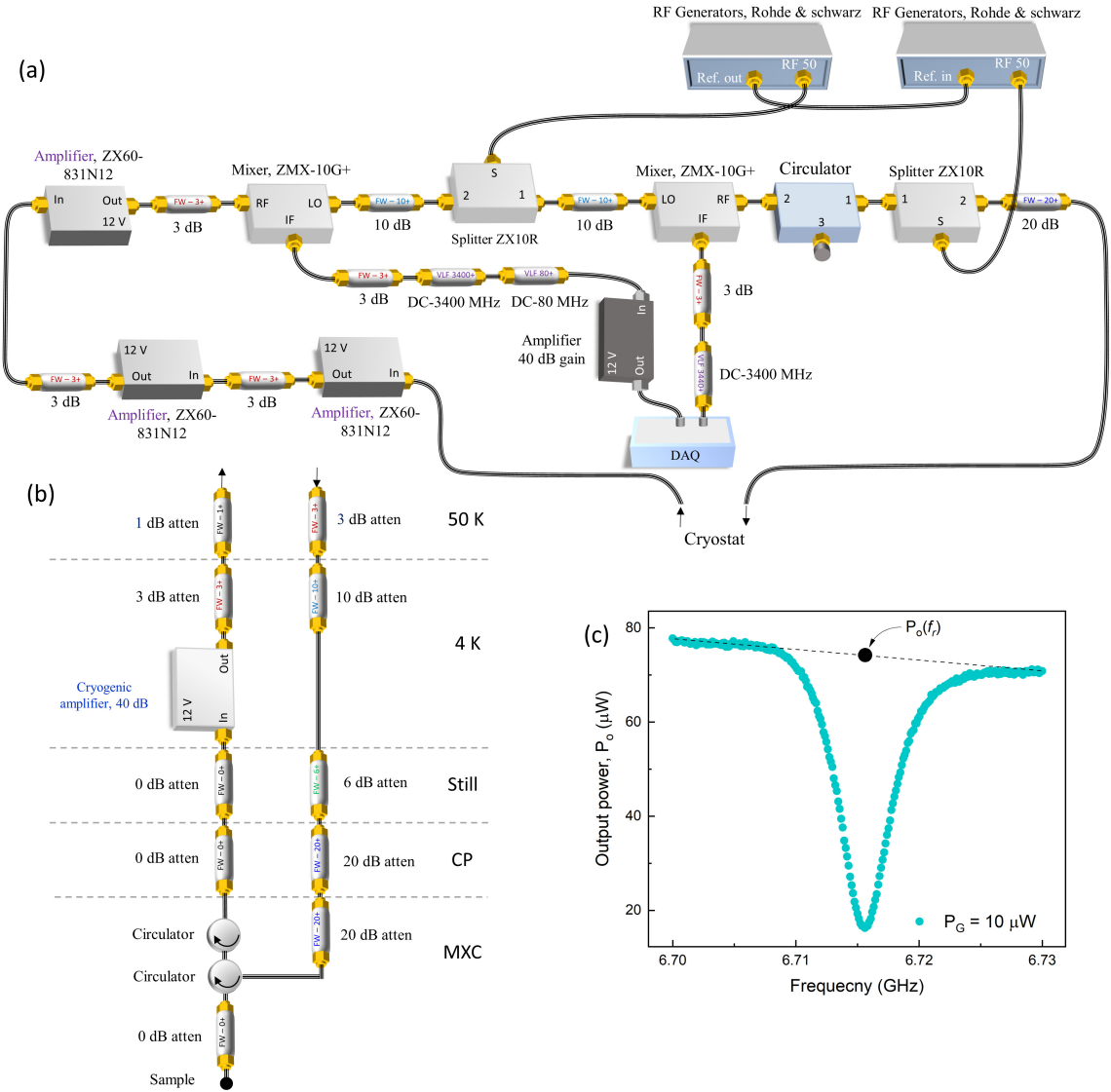}
\caption{\label{fig2s}The schematic of the RF heterodyne circuit arrangement showing (a) the room temperature board and (b) components inside the cryostat. (c): The RF reflected power as a function of frequency with an input power of -20 dBm. The background signal at the resonance is marked with a black dot.}
\end{figure}

Figures~\ref{fig2s}(a) and \ref{fig2s}(b) depict the schematic of our RF heterodyne scheme, which is used to measure the microwave response of the device. The room temperature board (Fig.~\ref{fig2s}(a) starts with an input signal from an RF generator, which is then split into two. Among the two split signals, one is used as a phase reference, which is mixed with a local oscillator (LO). Another part goes to the cryostat (Fig.~\ref{fig2s}(b), which is then fed into the device. The output signal is amplified at the 4K stage (+40.8 dB gain) and by three RT amplifiers (+61.5 dB gain), which is then mixed with the LO once again. The amplitude and phase responses are measured with a data acquisition (DAQ) system by determining the amplitudes and phases at the intermediate frequency of 1 kHz. In addition, the input and output lines consist of a few attenuators, two circulators, and low-pass filters, see Figs.~\ref{fig2s}(a) and \ref{fig2s}(b). Supplementary Table~\ref{tab1} shows the components used in the heterodyne board with their attenuation/amplification in dB as specified in the calibration data sheets. Here, the cable loss for the input and output lines is an unknown parameter, which we denote with $'x'$. We calculate the total attenuation used between the RF generator and the measuring probe is (33.67 - 2$x$). Here, we consider the cable loss on the input and output lines to be the same because of their nearly equal length. To determine the cable loss, we measure the RF reflected power $P_o$ at the data acquisition board (DAQ) as a function of frequency Fig.~\ref{fig2s}(c) and estimate the background reflected power $P_o(f_r)$ = 74~$\mu$W at the resonance point, marked with a black dot. The RF response is recorded with the input power at the generator set to $P_\text{G}$ = -20 dBm (= 10~$\mu$W). Since the reflection coefficient is expected to be the unity when there is no resonance, we equate $(33.67-2x) = 10\log({P_\text{o}/P_\text{G}})$. This yields us, cable loss x = -12.5 dB and the total attenuator on the input line -99.7 dB.
\begin{table*}
\caption{\label{tab1}Components used in the RF reflectrometry and their attenuator/amplification at 6.7 GHz.} 
\begin{ruledtabular}
\begin{tabular}{lclc}
Input line& (dB)& Output line& (dB)\\
\hline
Output inter-connect cable, Minicircuits 141-36SM+ & -1.08 & 2 $\times$ Cryogenic circulator, LNF-CIISC4 8A & -0.34  \\
 Power splitter, Minicircuits ZX10R-14-S+ & -7.12 & Cryogenic amplifier, LNF-LNC4 8C
 & +40.8 \\
 Room  temperature board attenuator &  -20& Attenuators in the cryostat
 & -4   \\
 Attenuators in the cryostat & -59 & 3 $\times$ RT amplifier, Minicircuits ZX60-83LN12+
  &  +61.5\\
 Cable loss & -x & 3 $\times$ Attenuators at RT & -9\\
 &&Mixer, Minicircuits ZMX-10G+, Conversion loss &-4.65\\
 &&IF attenuator&-3\\
 &&IF amplifier, Femto DLPVA &+40\\
 &&LP filter attenuation, VLF 3400+, VLF 80+ &-0.32\\
&&Input inter-connect cable, Minicircuits 141-2SM+&-0.12\\
&& Cable loss&-x\\
\hline
Total &(-87.2-x) &&(+120.87 –x) \\
 \end{tabular}
\end{ruledtabular}
\end{table*}

\section{External and internal quantum efficiency}

We calculated the photon-to-electron conversion efficiency $\eta_\text{PD}$ as the ratio of the photoelectron tunnel rate ($I_\text{SD}/e$) across the source-drain contacts to the photon flux ($P_\text{in}/hf_\text{r}$) irradiated on the resonator input from an external source. See the main text. Therefore, $\eta_\text{PD}$ considers the entire process of photon-assisted tunneling, which includes coupling losses, internal cavity losses, and the efficiency of the quantum transition.  Consequently, $\eta_\text{PD}$ corresponds to the external quantum efficiency (EQE) of the device. 

On the contrary, the internal quantum efficiency (IQE) refers to the efficiency of converting absorbed cavity photons into the photon-assisted tunneling events. In other words, IQE represents the ratio of photon-assisted tunnel rates ($I_\text{SD}/e$) to the photon absorption rate ($R_\text{abs} = \kappa_\text{DQD}n_\text{r}$). Here, $\kappa_\text{DQD}$ denotes the cavity photon dissipation rate due to the DQD absorption, and $n_r$ = $2Z_0C_\text{r}Q_\text{L}^2P_\text{in}/Q_\text{ex}hf_\text{r}$ the number of photons stored in the resonator~\cite{haldar2022}. Here, $C_\text{r}$ = 540~fF is the capacitance of the resonator line, $Z_0 = 58~\Omega$ is the impedance of the resonator line, and $Q_\text{L}$ ($Q_\text{ex}$) denotes the loaded (external) quality factor of the resonator. At $P_\text{in}$ = 1~fW, we calculate IQE = $I_\text{SD}/e\kappa_\text{DQD}n_\text{r}$ = 47\:\% (28\:\%) at the positive (negative) detuning condition. With 100~fW input power, the IQE of the DQD further decreases to 3.7\:\% (0.9\:\%) at +$\delta_r$ ($-\delta_r$).

\section{Theory}

We describe the driven DQD-resonator system using a standard Jaynes-Cummings Hamiltonian in the rotating frame of the incoming microwave radiation with frequency $\omega_{\text{l}}$
\begin{equation}\label{hst}
    H = \Delta_{\text{d}}\frac{\sigma_z}{2} + \Delta_{\text{r}} a^{\dagger}a + g\left(a^{\dagger} \sigma_- + a \sigma_+ \right) 
    + \sqrt{\kappa_{\text{c}} \dot{N}} \left( a^{\dagger} + a \right).
\end{equation}
Here, $\Delta_{\text{d}} = E_{\text{d}} -\omega_{\text{l}}$, where $E_{\text{d}} = \sqrt{\delta^2+4t^2}$ is the DQD level splitting, and $\Delta_{\text{r}} = E_{\text{r}}-\omega_{\text{l}}$, where $E_{\text{r}}$ is the resonator photon energy. The three DQD states are the ground state $\ket{g}$, the excited state $\ket{e}$ and the empty state $\ket{0}$, where no electron resides on the DQD. The Pauli spin matrices are defined in the subspace of ground and excited state, $\sigma_z=\ketbra{e}{e}-\ketbra{g}{g}$ and $\sigma_+=(\sigma_-)^{\dagger}=\ketbra{e}{g}$. The energy of the empty state is set to zero. The operators $a^{\dagger}$ ($a$) create (destroy) a photon in the resonator. The DQD-resonator coupling is given by $g=g_0\sin(\theta)$, where $g_0$ is the bare coupling and the mixing angle $\theta$ is given by $\cos(\theta)=-\delta/E_{\text{d}}$. Lastly, the drive amplitude is a product between the rate of incident photons $\dot{N}$ and the coupling to the input port $\kappa_{\text{c}}$. 

The dynamics of the DQD-resonator system, including the coupling of the DQD to source and drain contacts, is described by a Lindblad master equation  
\begin{equation}\label{lme}
\begin{split}
\partial_t \rho(t)=&-i[H,\rho(t)] + \Gamma_{g0} \mathcal{D}[s_g^{\dagger}]\rho(t) + \Gamma_{0g} \mathcal{D}[s_g]\rho(t)+ \Gamma_{0e} \mathcal{D}[s_e]\rho(t)+ \Gamma_{e0} \mathcal{D}[s_e^\dagger]\rho(t) + \gamma_{\phi}\mathcal{D}[\sigma_z]\rho(t) +\gamma_- \mathcal{D}[\sigma_-]\rho(t)\\[0.25em]&
+\kappa \mathcal{D}[a]\rho(t), 
\end{split}
\end{equation}
with the Lindblad superoperator $\mathcal{D}[x]\rho(t)= x\rho(t) x^{\dagger} - \frac{1}{2}\{x^{\dagger}x,\rho(t) \}$, where $\rho(t)$ is the density matrix of the DQD-resonator system. Here, $\Gamma_{g0}$ ($\Gamma_{0g}$) denotes the rate for an electron tunneling into (out of) the ground state, and similarly $\Gamma_{e0}$ ($\Gamma_{0e}$) for the excited state. The rates are given by
\begin{equation}
\begin{split}
    &\Gamma_{g0}=\Gamma_{g\text{L}}+\Gamma_{g\text{R}}, \qquad \Gamma_{0g}=\Gamma_{\text{L}g}+\Gamma_{\text{R}g}, \\
    &\Gamma_{e0}=\Gamma_{e\text{L}}+\Gamma_{e\text{R}}, \hspace{0.75cm} \Gamma_{0e}=\Gamma_{\text{L}e}+\Gamma_{\text{R}e},
\end{split}
\end{equation}
where
\begin{equation}
\begin{split}
    &\Gamma_{g\text{L}}= \Gamma_{\text{L}}f_{\text{L}}(E_g)\cos^2(\theta/2), \hspace{2cm} \Gamma_{g\text{R}}=\Gamma_{\text{R}}f_{\text{R}}(E_g)\sin^2(\theta/2), \\
    &\Gamma_{\text{L}g}= \Gamma_{\text{L}}[1-f_{\text{L}}(E_g)]\cos^2(\theta/2), \hspace{1.2cm} \Gamma_{\text{R}g}=\Gamma_{\text{R}}[1-f_{\text{R}}(E_g)]\sin^2(\theta/2), \\
    &\Gamma_{e\text{L}}= \Gamma_{\text{L}}f_{\text{L}}(E_e)\sin^2(\theta/2), \hspace{2.07cm} \Gamma_{e\text{R}}=\Gamma_{\text{R}}f_{\text{R}}(E_e)\cos^2(\theta/2), \\    
    &\Gamma_{\text{L}e}= \Gamma_{\text{L}}[1-f_{\text{L}}(E_e)]\sin^2(\theta/2), \hspace{1.27cm} \Gamma_{\text{R}e}=\Gamma_{\text{R}}[1-f_{\text{R}}(E_e)]\cos^2(\theta/2).
\end{split}
\end{equation}
Here, $\Gamma_{\text{L}}$ $(\Gamma_{\text{R}})$  is the rate for electron tunneling events into or out of the left (right) dot and $f_{\text{L,R}}(E)$ are the Fermi-Dirac distributions given by $f_{\text{L}}(E)=1/\{1+\exp[E/(k_BT)]\}$ and $f_{\text{R}}(E)=1/\{1+\exp[(E-V_b)/(k_BT)]\}$ with $V_b$ denoting the applied bias across source and drain.
The operators $s_g=\ketbra{0}{g}$ and $s_e=\ketbra{0}{e}$ remove an electron in the ground and excited state, respectively. The rate $\gamma_{\phi}$ denotes pure dephasing of the DQD (caused by voltage fluctuations in the electromagnetic environment), $\gamma_-$ denotes the internal relaxation rate (through coupling of electrons to phonons in the solid state environment) and $\kappa$ denotes the total photonic loss rate. 

The electrical current flowing from source to drain is given by 
\begin{equation}\label{cur2}
    I_{\text{SD}}/e=\Gamma_{\text{R}e}\mel{e}{\rho}{e}+\Gamma_{\text{R}g}\mel{g}{\rho}{g}-(\Gamma_{g\text{R}}+\Gamma_{e\text{R}})\mel{0}{\rho}{0}. 
\end{equation}
The current at detunings $\delta=\pm\delta_r$ is denoted by $I_{\text{SD}}^{\pm}$ in the main text.
\subsection{Low-drive limit}

We first consider the case where zero bias is applied across the DQD. Since the temperature of the leads is negligible compared to the DQD level spacing $k_BT\ll E_{\text{d}}$, we can in this case put $f_{\text{L,R}}(E_g)\to 1$ and $f_{\text{L,R}}(E_e)\to 0$.

We further work in the low-drive limit, where $\dot{N} \to 0$. In this limit, the master equation can be expanded to lowest order in $\sqrt{\kappa_{\text{c}} \dot{N}}$, which results in the current
\begin{equation}\label{cur}
    I_{\text{SD}}/e= \Gamma_{\text{LR}}-\Gamma_{\text{RL}},
\end{equation}
with
\begin{equation}\label{l0perf1}
    \Gamma_{\alpha \beta}=\frac{16\kappa_{\text{c}}\dot{N}g^2[4\kappa\Delta_\text{d}(\Delta_\text{d}-\Delta_\text{r})+(\tilde{\Gamma}+\kappa)(4g^2+\tilde{\Gamma}\kappa)]\Gamma_{g\alpha}\Gamma_{\beta e}}{\Gamma_{g0}[4g^2(\Gamma_{0e}+\gamma_-+\kappa)+\kappa(\Gamma_{0e}+\gamma_-)(\tilde{\Gamma}+\kappa)][(\tilde{\Gamma}^2+4\Delta_{\text{d}}^2)(\kappa^2+4\Delta_{\text{r}}^2)+8g^2(\tilde{\Gamma}\kappa-4\Delta_{\text{d}}\Delta_{\text{r}})+16g^4]},
\end{equation}
where $\tilde{\Gamma}\equiv\Gamma_{0e}+\gamma_-+2\gamma_{\phi}$ is the total decoherence rate and $\alpha,\beta\in\{\text{L, R}\}$. We obtain the expression in Eq.~(3) in the main text from Eq.~\eqref{l0perf1} by setting $\Delta_{\text{d}}=\Delta_{\text{r}}=0$ and further assuming that $\Gamma_{0e} \gg \kappa$. This equation is then used to produce the solid lines in Fig.~2(c). Note also that the rate of incident photons is related to the input power by $\dot{N}=P_{\text{in}}/hf_r$. 
The photo-detection efficiency in the low-drive limit is defined as
\begin{equation}
    \eta_{\text{PD}}=\frac{I_{\text{SD}}/e}{\dot{N}},
\end{equation}
At detunings $\delta=\pm\delta_r$, the efficiency is denoted by $\eta_{\text{PD}}^\pm$ in the main text, as shown in Fig~2d. The low-drive photo-detection efficiency is independent on the drive strength, as is clear by dividing the current in Eq.~\eqref{cur} by $\dot{N}$.

\subsection{Large-drive limit}
In the large-drive limit, the quantum fluctuations of the resonator photon state are small and we may neglect back-action of the DQD. We can then replace the photonic annihilation operator $a$ with a c-number given by $-2if/\kappa$, equal to the value of $\expval{a}$ for $g=0$. This results in the current 
\begin{equation}\label{eq:l}
    I_{\text{SD}}/e = \frac{16\kappa_{\text{c}}\dot{N}g^2(\Gamma_{g\text{L}}\Gamma_{\text{R}e}-\Gamma_{g\text{R}}\Gamma_{\text{L}e})}{16\kappa_{\text{c}}\dot{N}g^2(\Gamma_{0e}+2\Gamma_{g0})+\Gamma_{g0}\tilde{\Gamma}\kappa^2(\Gamma_{0e}+\gamma_-)},
\end{equation}
which is the equation used to produce the dashed lines in Fig.~2(c). The expression for the saturation current $I^{*}_{\text{SD}} $ in the main text is then found by taking the limit $\dot{N} \to \infty.$
\subsection{Reflectance}
To evaluate the reflection coefficient $R$, we use input-output theory, which results in the following Heisenberg equations of motion in the low-drive regime 
\begin{equation}\label{eq:io}
\begin{split}
    \partial_t \expval{a}(t)&=-\frac{\kappa}{2}\expval{a}(t)-i\Delta \expval{a}(t)-ig\expval{\sigma_-}(t)+\sqrt{\kappa_{\text{c}}}\expval{b_{\text{in}}}(t), \\
    \partial_t \expval{\sigma_-}(t) &= -\frac{\tilde{\Gamma}}{2}  \expval{\sigma_-}(t)-i\Delta \expval{\sigma_-}(t)-ig \expval{a}(t),
\end{split}
\end{equation}
where we assume a resonant DQD-resonator coupling, $E_{\text{d}}=E_{\text{r}}$, and hence $ \Delta_{\text{d}}=\Delta_{\text{r}}\equiv\Delta$. Taking the Fourier transform of Eqs. \eqref{eq:io} leads to 
\begin{equation}\label{ao}
    \expval{a}(\omega)= \frac{\sqrt{\kappa_{\text{c}}}\left[\frac{\tilde{\Gamma}}{2}-i(\omega-\Delta)\right]}{\left[\frac{\tilde{\Gamma}}{2}-i(\omega-\Delta)\right]\left[\frac{\kappa}{2}-i(\omega-\Delta)\right]+g^2}\expval{b_{\text{in}}}(\omega). 
\end{equation}
We note that since we are working frame rotating around the frequency of the drive, we need to evaluate the zero frequency component of Eq.~\eqref{ao}. We further need the standard input-output relations,
\begin{equation}
    \expval{b_{\text{in}}}(\omega) + \expval{b_{\text{out}}}(\omega) = \sqrt{\kappa_{\text{c}}}\expval{a}(\omega).
\end{equation}
The reflection coefficient is given by $R=\abs{r}^2$ with the reflection amplitude being 
\begin{equation}
    r = \frac{ \expval{b_{\text{out}}}(\omega=0)}{\expval{b_{\text{in}}}(\omega=0)} = \frac{\kappa_{\text{c}}\left[\frac{\tilde{\Gamma}}{2}-i(\omega_{\text{l}}-E_{\text{r}})\right]}{\left[\frac{\tilde{\Gamma}}{2}-i(\omega_{\text{l}}-E_{\text{r}})\right]\left[\frac{\kappa}{2}-i(\omega_{\text{l}}-E_{\text{r}})\right]+g^2}-1.
\end{equation}
If we further assume that $\tilde{\Gamma}\gg\kappa$, the reflection coefficient simplifies to
\begin{equation}\label{refl}
    R=\frac{\left(\frac{\kappa}{2}+\frac{\kappa_{\text{DQD}}}{2}-\kappa_{\text{c}}\right)^2+(\omega_{\text{l}}-E_{\text{r}})^2}{\left(\frac{\kappa}{2}+\frac{\kappa_{\text{DQD}}}{2}\right)^2+(\omega_{\text{l}}-E_{\text{r}})^2},
\end{equation}
where $\kappa_{\text{DQD}}=4g^2/\tilde{\Gamma}$. Eq.~\eqref{refl} was used to produce the solid lines in Fig.~2(a). In the Coulomb blockade regime ($\delta \gg \delta_r$), the DQD and the resonator are effectively decoupled, which can be modeled by setting $g=0$. 

\subsection{Finite voltage bias}
For a finite voltage bias $V_b$ applied between source and drain, the currents $I_{\text{SD}}^{\pm}$ (solid lines in Fig.~3(a) are found by solving the master equation \eqref{lme} numerically and using Eq.~\eqref{cur2}. For the 100 fW case (solid lines in Fig.~3(b), we use the large-drive approximation introduced above, solve the master equation numerically and use Eq.~\eqref{cur2}. The numerical computations are done using QuTip \cite{JOHANSSON20121760}.

\subsection{Validity of the master equation}
The master equation in Eq.~\eqref{lme} has a restricted regime of validity due to the approximations used in deriving it. First, it is a \textit{local} master equation in the sense that the environment couples to the cavity and the DQD through independent jump terms. This is justified as long as the coupling between the DQD and the cavity is sufficiently weak. More explicitly, for weak power, we expect the local approach to be valid as long as \cite{Potts_2021}
\begin{equation}
\label{eq:local}
    f_\alpha(E_e\pm g)\simeq f_\alpha(E_e),\hspace{1cm}  f_\alpha(E_g\pm g)\simeq f_\alpha(E_g),\hspace{1cm} g\ll E_{\rm r},
\end{equation}
where $\alpha =$ L, R. At strong input power, the coupling is enhanced by the average displacement in the cavity and we need to replace the Jaynes-Cummings coupling in the last expression as $g\rightarrow g|\langle a\rangle |$.

In case the local approach is justified, the Born-Markov approximations are justified as long as the reservoir occupations do not vary by much across the broadened energy levels
\begin{equation}
\label{eq:BM}
  f_\alpha(E_e\pm (\Gamma_\mathrm{L}+\Gamma_\mathrm{R}))\simeq f_\alpha(E_e),\hspace{1cm}  f_\alpha(E_g\pm (\Gamma_\mathrm{L}+\Gamma_\mathrm{R}))\simeq f_\alpha(E_g),  \hspace{1cm} \kappa\ll E_{\rm r}
\end{equation}
where, for simplicity, we approximated the broadening of the DQD levels by $\Gamma_\mathrm{L}+\Gamma_\mathrm{R}$.

We note that both Eqs.~\eqref{eq:local} and \eqref{eq:BM} can be satisfied either by sufficiently high temperatures or by DQD energies that are sufficiently far away from the chemical potentials. Additionally, the reservoir of the cavity needs to have a negligible occupation at all relevant energies.

Furthermore, the jump terms corresponding to the DQD are obtained using a rotating wave approximation which is justified as long as
\begin{equation}
    \label{eq:RWA}
    E_{\rm d} \gg \Gamma_\mathrm{L}+\Gamma_\mathrm{R}.
\end{equation}

Finally, the Jaynes-Cummings Hamiltonian neglects counter-rotating terms of the form of $a^\dagger\sigma_+$. These terms can be neglected as long as the effective coupling is small compared to the cavity frequency
\begin{equation}
    \label{eq:counterrot}
    E_{\rm r}+E_{\rm d}\gg g|\langle a\rangle|\simeq g\sqrt{\frac{\kappa_\mathrm{c}\dot{N}}{\kappa^2+(\Delta_r/2)^2}},
\end{equation}
where the last expression holds in the large-drive limit. We note that Eq.~\eqref{eq:counterrot} is very similar to the last constraint that justifies the local approach in Eq.~\eqref{eq:local}. Indeed, it is known that at strong coupling (or strong drives), the local approach breaks down \cite{forn-diaz_2019}.

\section{Effect of Temperature on the power harvesting efficiency}
\begin{figure}[h]
\includegraphics[width=6.7in]{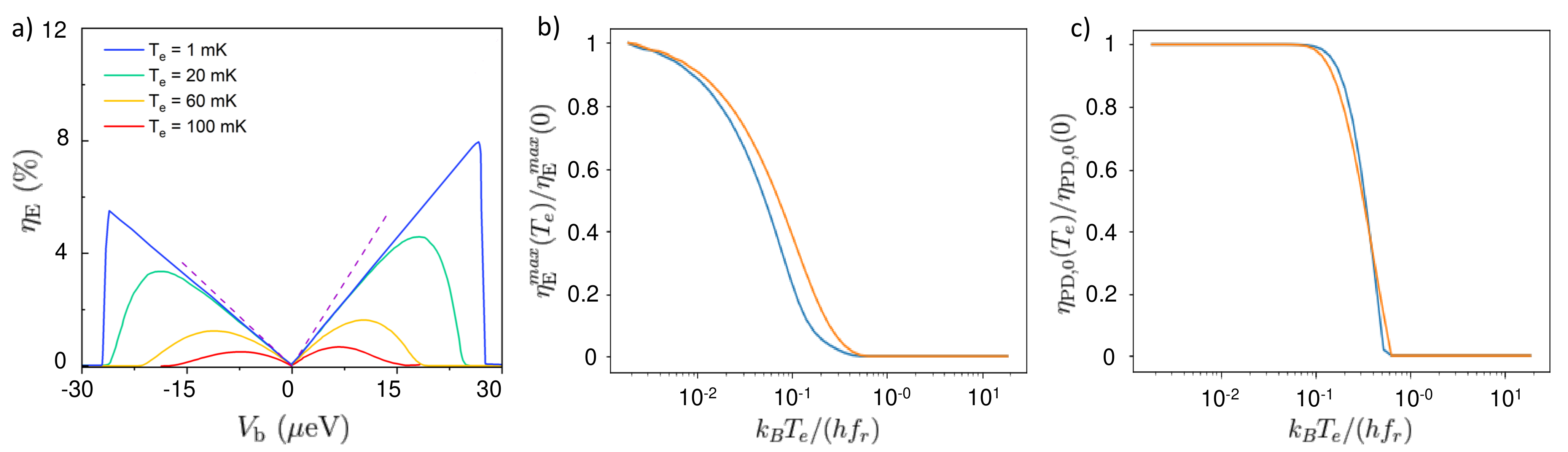}
\caption{\label{fig3s}(a) Power harvesting efficiency $\eta_E$ calculated theoretically for different temperatures with $P_\text{in}$ = 1~fW. The dashed purple line shows $\eta_E^\pm=\eta_\text{PD}^\pm eV_b/hf_{\text{r}}$ for the low-bias limit. With decreasing temperature, the stopping potential $V_b^s$ moves towards the single photon energy ($hf_r=27.7~\mu$eV) and the maximum power harvesting efficiency $\eta_E$ approaches the photon-to-electron conversion efficiency $\eta_\text{PD}$. In panels (b) and (c), the maximum power harvesting efficiency $\eta_\text E^{max,\pm}$ and the photodetection efficiency at low voltage bias $\eta_{\text{PD},0}^{\pm}$ are plotted as a function of temperature, normalized with their respective $T_e=0$ values. In both panels the level detuning is $+$ (blue line) and $-$ (red line)}.
\label{SIfig3}
\end{figure}

To investigate the effect of temperature on the power harvesting efficiency we numerically calculate $\eta_E$ as a function of $V_b$ for different electronic temperatures $T_e=1, 20, 60$ and $100$~mK of the reservoir, where we note that the electronic temperature in the experiment is $60$~mK. All other parameter values are considered to be the same as in Fig.~3(e) in the main text for this analysis. Supplementary Figure~\ref{fig3s}(a) shows that  $\eta_E(V_b)$ is monotonically suppressed by increasing temperature, such that $|V_b^{\text{m}}| , \eta_E^\text{max}$ and the stopping-potential $|V_b^s|$, when the microwave activated current becomes equal to the counter current due to the applied bias, are all suppressed with increasing $T_e$, for both positive and negative voltage bias. We note that for $T_e\rightarrow$ 0, the maximum power harvesting efficiency approaches the zero-bias photon-to-electron conversion efficiency $\eta_\text{PD}$ at $|V_b^{\text{m}}| \approx hf_r$.
In panels (b) and (c) the suppression of the photodetection efficiency at low voltage bias and the maximum power harvesting efficiency are plotted as a function of the temperature $T_e$. These panels clearly illustrate that the power harvesting efficiency is more strongly suppressed by increasing temperature compared to the photodetection efficiency. This observation is further discussed and explained below. 

\subsection{Basic meanfield model}
To get a qualitative, analytical understanding of the effect of temperature on the photodetection and power harvesting efficiencies, we consider a particularly simple parameter regime where the photodetection efficiency, $\eta_\text{PD}$, reaches unity, as discussed in Refs. \cite{wong2017,khan2021}. Here we take identical tunnel barriers, $\Gamma_L=\Gamma_R=\Gamma$ and zero dephasing, $\gamma_{\phi}=0$, relaxation, $\gamma_-=0$ and internal photon losses, $\kappa_{\text{int}}=0$. Moreover, $\kappa=4g^2/\Gamma$, with $g=g_0\sin(\theta)$ and $\kappa=\kappa_c$, the microwave drive is in the weak, linear, regime and the directivity is close to unity, corresponding to a mixing angle $\theta \ll 1$. However, in contrast to ideal photodetector operation, which requires a reservoir electron temperature energy $k_BT_e$ well below the photon energy $hf_r$, we here consider arbitrary temperature. 

\begin{figure}[h]
  \centering
{\includegraphics[width=6.7in]{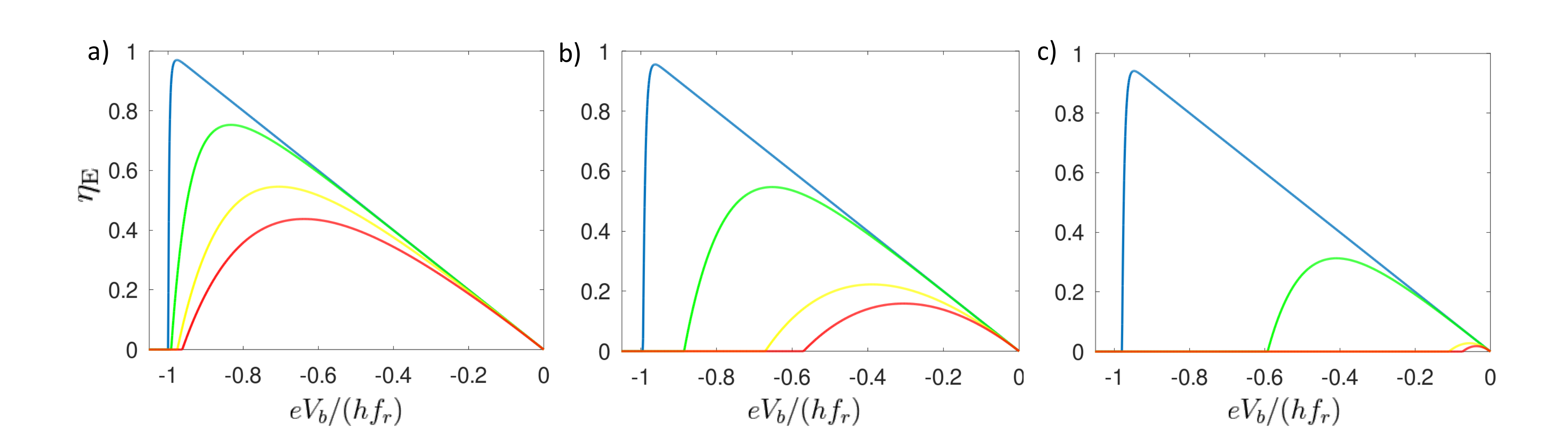}}  
\caption{Power harvesting efficiency as a function of applied bias $eV_b/hf_r$, for different temperatures $T_e=1$~mK (blue), $20$~mK (red), $60$~mK (yellow) and $100$~mK (purple), same as in Fig.~\ref{SIfig3}. The parameter $\zeta=1$ (a) , $10$ (b) and $100$ (c).}
\label{fig4s}
\end{figure}

To harvest power we consider a voltage bias $V_b$, symmetrically applied across the double dot. In addition, we treat the equations of motion for the system density matrix in the mean field approximation, found in Refs. \cite{khan2021,zenelaj2022} to give accurate results when compared to full numerics. Under these conditions, we can express the photodetection efficiency for positive ($+$) level detuning as
\begin{equation}
\eta_{\text{PD}}=\zeta\frac{\cosh\left[\frac{hf_r + eV_b}{4k_BT_e}\right] \sinh\left[\frac{eV_b}{2k_BT_e}\right]}{2\cosh\left[\frac{hf_r - eV_b}{4k_BT_e}\right]\left(1+2\cosh\left[\frac{hf_r + eV_b}{2k_BT_e}\right]\right)}+\frac{8\left(\sinh\left[\frac{hf_r + eV_b}{2k_BT_e}\right] + \sinh\left[\frac{hf_r + eV_b}{k_BT_e}\right]\right)}{\left(1 + 2 e^{(hf_r+eV_b)/(2k_BT_e)}\right)^2},
\label{PDeq}
\end{equation} 
where the dimensionless parameter $\zeta=\theta^2\Gamma hf_r/P_{\text{in}}$. Here the first term corresponds to the counter current due to applied voltage while the second corresponds to the photocurrent. We note that under the conditions $\zeta \ll 1$, $hf_r-eV_b \gg k_BT_e$ we recover $\eta_{\text{PD}}=1$, ideal photodetection. As pointed out in the main text, the power harvesting efficiency is given by
\begin{equation}
\eta_E=\eta_{\text{PD}}\frac{e|V_b|}{hf_r}.
\label{Eeff}
\end{equation}
In Fig. \ref{fig4s} we plot $\eta_E$ as a function of $eV_b/(hf_r)$ for different temperatures and parameters $\zeta$. From the plots it is clear that the power harvesting efficiency is decreasing for increasing temperature and $\zeta$. 
To evaluate the maximum power harvesting efficiency as a function of voltage $V_b$, for a given $T$ and $\zeta$, we find the maximum of Eq.~(\ref{PDeq}) numerically. We note that a full numerical evaluation with $g_0 \ll \Gamma$  (not shown here) gives an agreement with Fig.~\ref{fig4s} which is qualitative for $\zeta \gtrsim 1$ and becomes quantitative for $\zeta \ll 1$.

\begin{figure}[h]
  \centering
 {\includegraphics[scale=0.5]{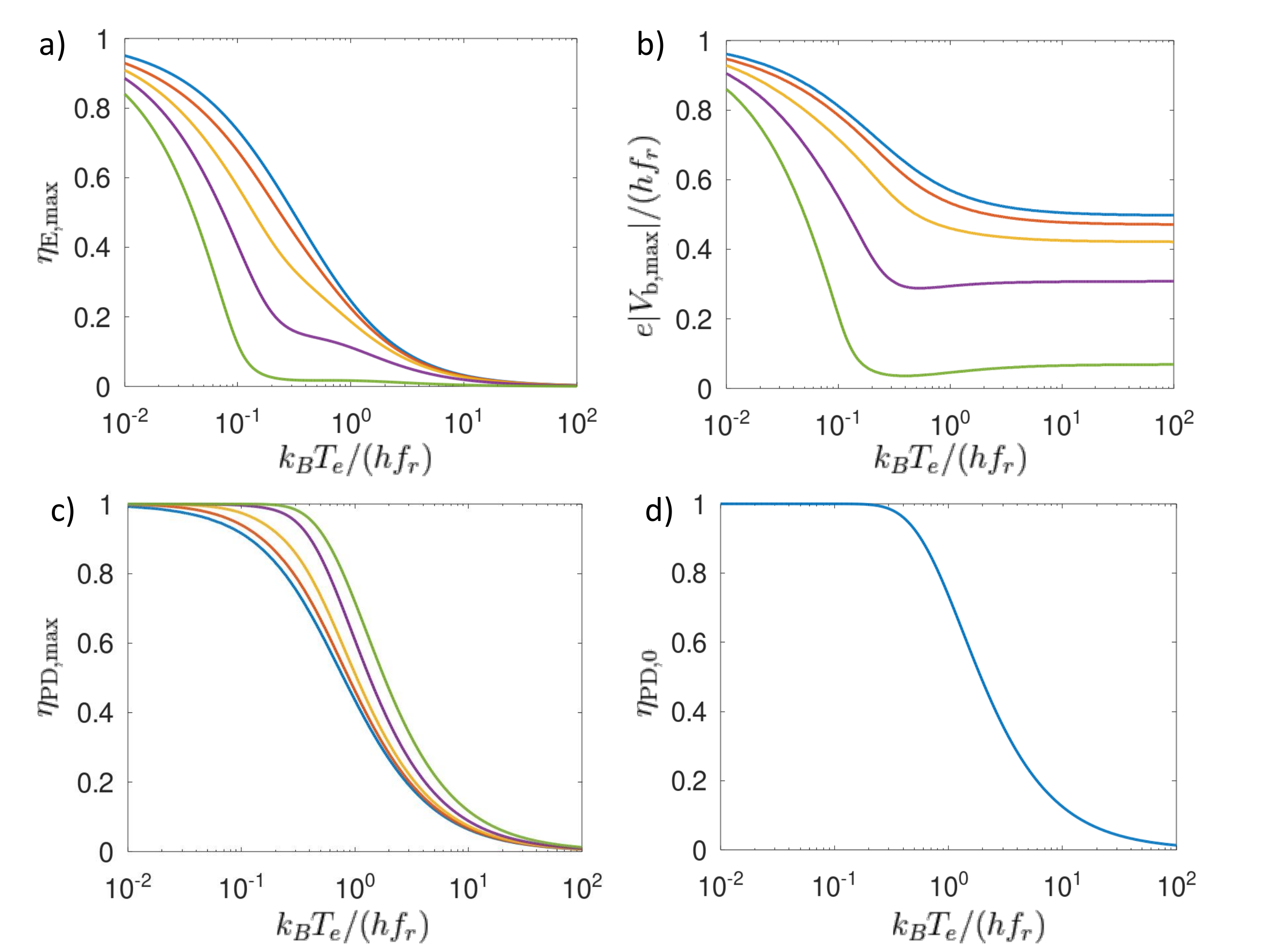}}  
  \caption{The maximum power harvesting efficiency $\eta_{E}^\text{max}$ (a), maximizing voltage bias $|V_{b}^\text{m}|/(hf_r)$ (b) and  the photodetector efficiency $\eta_\text{PD}^{max}$ (c) as a function of temperature $k_BT_e/(hf_r)$, for different values of $\zeta$. In panels a, b and c, $\zeta=0.1$ (blue), $1$ (red), $3$ (yellow), $10$ (purple) and $100$ (green). In panel (d), the photodetector efficiency at zero voltage bias (independent on $\zeta$), as a function of temperature is shown.}
\label{fig5s}
\end{figure}

The maximum power harvesting efficiency, $\eta_{E}^\text{max}$, together with the maximizing bias voltage $V_{b}^\text{m}$, the photodetector efficiency at $V_{b}^\text{m}$, denoted $\eta_\text{PD}^\text{max}$, and the photodetector efficiency at zero voltage bias, $\eta_{\text{PD,0}}$ are plotted as a function of temperature in Fig.~\ref{fig5s} for a set of different $\zeta$. 
From the plots in Fig.~\ref{fig5s} it is clear that the maximum power harvesting efficiency is more strongly suppressed due to increasing temperature, compared to the photodetector efficiency at $V_b=V_\text{b}^\text{m}$ (as well as $V_b=0$). The key reason, evident from Eq.~(\ref{Eeff}), is that $\eta_\text{E}^\text{max}$ is the product of $\eta_\text{PD}^\text{max}$ and the ratio $e|V_\text{b}^{m}|/(hf_r)$. Since the voltage bias that maximizes the power harvesting efficiency is always smaller than the resonance energy, or DQD level splitting, $hf_r$, we have that $\eta_\text{E}^\text{max} \leq \eta_\text{PD}^\text{max}$. 

This inequality can be quantified by considering the simplest possible limit, $\zeta \rightarrow 0$, where the counter current can be neglected. The effect of a non-zero but small temperature $k_BT_e \ll hf_r$ can then be analytically evaluated, giving the maximizing voltage bias (keeping leading order terms)
\begin{equation}
eV_\text{b}^\text{m}=-hf_r+2k_BT_e\ln\left[\frac{1}{2}\sqrt{\frac{hf_r}{k_BT_e}}+\frac{9}{4}\right]
\end{equation}
and the efficiencies
\begin{equation}
\eta_\text{PD}^\text{max}=1-\frac{k_BT_e}{hf_r}, \hspace{0.5 cm} \eta_\text{E}^\text{max}=\left(1-\frac{k_BT_e}{hf_r}\right)\left(1-2\frac{k_BT_e}{hf_r}\ln\left[\frac{1}{2}\sqrt{\frac{hf_r}{k_BT_e}}+\frac{9}{4}\right]\right) \leq \eta_\text{PD}^\text{max}.
\end{equation}
In the opposite limit $k_BT_e \gg hf_r$ we find the maximizing voltage bias
\begin{equation}
eV_\text{b}^\text{m}=-\frac{hf_r}{2}\left(1+ \frac{hf_r}{6k_BT_e}\right)
\end{equation}
and the efficiencies
\begin{equation}
\eta_\text{PD}^\text{max}=\frac{2hf_r}{3k_BT_e}, \hspace{0.5 cm} \eta_\text{E}^\text{max}=\frac{hf_r}{3k_BT_e}\left(1+ \frac{hf_r}{6k_BT_e}\right)  \leq \eta_\text{PD}^\text{max}.
\end{equation}
\subsection{Qualitative comparison to full numerics and experiment}
It is instructive to compare the results of the basic meanfield model to the full numerics in Fig.~\ref{fig3s}, which describes well the experimental results. Despite the fact that the underlying assumptions and parameters are different, several qualitative results are similar:\\
$\bullet$ The overall effect of increasing temperature is to monotonically suppress both $\eta_{\text{PD}}$ and $\eta_{\text{E}}$, for all bias voltages. Moreover, the voltage that maximizes $\eta_{\text{E}}$ is always smaller than (or equal to) the photon energy,  $e|V_\text{b}^\text{m}|\leq hf_r$, and decreases with increasing temperature.\\
$\bullet$ The maximum energy harvesting efficiency $\eta_\text{E}^\text{max}$ is always smaller than the photodetection efficiency, $\eta_\text{PD}^\text{max}$, at maximizing voltage bias $eV_\text{b}^\text{m}$. This follows directly from $\eta_\text{E}^\text{max}=\eta_\text{PD}^\text{max}e|V_\text{b}^\text{m}|/(hf_r)\leq \eta_\text{PD}^\text{max}$.\\
$\bullet$ By comparing the temperature dependence of $\eta_\text{E}^\text{max}$ and $\eta_{\text{PD},0}$ evaluated within the meanfield model, Figs.~\ref{fig4s}(a) and (d), with the full numerics, Fig.~\ref{fig3s}(b), (c), we see that (up to a constant prefactor) the mean field model captures well the suppression of the efficiencies with increasing $T_e$. \\
$\bullet$ Moreover, the bias voltage dependence of $\eta_{\text{E}}$, in Fig.~\ref{fig3s}(a), is best described (up to a constant prefactor) by the mean field model for a value of the parameter $\zeta \approx 10$. Since the mean field model considers a symmetric system with $\Gamma=\Gamma_L=\Gamma_R$, one can not directly determine a value for $\zeta$ from the full numerics, where we have $\Gamma_R\gg \Gamma_L$. However, to get a rough estimate one can put $\Gamma=(\Gamma_R+\Gamma_L)/2$, giving a value $\zeta=2.4$, of the same order of magnitude as $10$.\\
$\bullet$ By analyzing the two components of the photodetector efficiency expression in Eq.~(\ref{PDeq}), we see that for $\zeta\approx 10$ and larger the counter current due to the applied bias starts to be of comparable magnitude (but opposite in sign) to the photogenerated current. Given the good qualitative, and in some aspects quantitative, agreement with the full numerics, and hence the experiment, from the investigations of the meanfield model we thus conclude that the counter current is relevant for determining the efficiencies in the experiment. 

\bibliography{master}